\documentclass{IEEEtran}
\pdfoutput=1
\usepackage{mathrsfs,amsmath,amssymb,mathtools,amsthm}
\usepackage{textcomp}
\usepackage{graphicx,multirow}
\usepackage{cite}
\usepackage{footnote}
\usepackage{booktabs}
\usepackage{stfloats}
\usepackage{colortbl}
\usepackage[inline]{enumitem}

\DeclareMathOperator{\tr}{Tr}
\DeclareMathOperator{\fim}{\mathbf{FIM}}
\DeclareMathOperator{\crb}{\mathbf{CRLB}}

\newcommand{\bs}{\boldsymbol}
\newcommand{\revisioncolor}{black}

\begin{document}

\title{Uncertainty Principle in Distributed MIMO Radars}

\author{Seyed MohammadReza Hosseini, Afshin Isazadeh, Ali Noroozi, and Mohammad Ali Sebt
\thanks{
The authors are with the Department of Electrical Engineering, K. N. Toosi university of Technology, Tehran, Iran. (email:
s.m.hosseini@ee.kntu.ac.ir;
a\_isazadeh93@ee.kntu.ac.ir;
ali\_noroozi@ee.kntu.ac.ir;
sebt@kntu.ac.ir)}
}
\markboth{
    Hosseini, Isazadeh, Noroozi and Sebt: 
    Uncertainty Principle in Distributed MIMO Radars}{}
\IEEEpubid{}
\maketitle

\begin{abstract}
Radar uncertainty principle indicates that there is an inherent invariance in the product of the time-delay and Doppler-shift measurement accuracy and resolution which can be tuned by the waveform at transmitter. In this paper,  based on the radar uncertainty principle, a conceptual waveform design is proposed for a distributed multiple-input multiple-output (MIMO) radar system in order to improve the Cramer-Rao lower bound (CRLB) of the target position and velocity. 
To this end, a non-convex band constrained optimization problem is formulated, and a local and the global solution to the problem are obtained by sequential quadratic programming (SQP) and particle swarm algorithms, respectively.
Numerical results are also included to illustrate the effectiveness of the proposed mechanism on the CRLB of the target position and velocity.
By numerical results, it is also concluded that the global solution to the optimization problem is obtained at a vertex of the bounding box. 

\end{abstract}

\IEEEpeerreviewmaketitle

\bstctlcite{IEEEexample:BSTcontrol}
\section{Introduction}\label{Introduction}
    \IEEEPARstart{M}{ultiple-input multiple-output} (MIMO) radar systems with widely separated antennas have been introduced to enhance detection and estimation performance by utilizing spatial diversity \cite{fishler2006spatial,godrich2010target}.
    In MIMO radar systems with co-located antennas waveform diversity also allows that more parameters can be identified unambiguously
    \cite{li2007parameter,li2007mimo}.
    Waveform design in conventional single-antenna radar has been considered extensively in the literature based on the optimizing a performance measure subject to some practical and functional constraints 
    \cite{soltanalian2014designing,aubry2015new}.
    Furthermore, the waveform design can be done adaptively based on the estimated target and environment parameters 
    \cite{de2008code,wu2017cognitive}.
    Waveform design has also been extended to the co-located MIMO radars 
    \cite{li2008mimo,Cheng2018}.

    For direct localization in distributed MIMO radars,
    which the target is directly estimated from the received signals, an adaptive mechanism for optimal energy allocation at different transmit antennas is proposed to improve the compressive-sensing algorithm performance \cite{gogineni2011target}. Therefore, there is a design parameter at the signal domain which can be interpreted as the waveform design in such a method.
    In indirect methods such as \cite{yang2016improved,amiri2017efficient,Amiri2017},
    which the target position and velocity are estimated based on the extracted measurements (bi-static range (BR) and bi-static range rate (BRR)) from the received signals,
    the transmitters' waveforms are not involved directly in the measurement model. Consequently, the capability of the waveform design at transmitters is ignored in the literature for such methods.

    In this paper, considering that the time-delay (TD) and Doppler-shift (DS) measurements are available at receivers, 
    we perform a conceptual waveform design at transmitters by applying the \emph{radar uncertainty principle}.
    Under constant compression ratio, 
    radar uncertainty principle indicates that there is a trade-off between the accuracy (and resolution) of the time-delay and Doppler-shift measurements which can be exchanged by the transmitter's waveform \cite{Vakmann1968}. Therefore, we strike a balance between the accuracy of TD and DS measurements for each transmitter in order to improve the Cramer-Rao lower bound (CRLB).

    We will abbreviate transmitter as Tx and receiver as Rx.
    Vectors and matrices are denoted by boldface lower- and upper-case letters, respectively.
    The $k$th element of the vector $\bs{a}$ is represented by $[\bs{a}]_{k}^{}$, and $[\bs{A}]_{k,:}$ denotes the $k$th row of the matrix $\bs{A}$. 
    $\bs{1}_k$ denotes the ${k\times1}$ vector of one.
    $\bs{0}$ and $\mathbf{O}$ are the zero vector and zero matrix proper in size, respectively. 
    The identity matrix of size ${k\times k}$  is represented by $\bs{I}_k^{}$.
    The superscripts $^T$ and $^{-1}$ denote the transpose and inverse operators, respectively.
    $\|\bs{a}\|$ 
    stands for the Euclidean norm of the vector $\bs{a}$.

    The remainder of the paper is organized as follows. Section II is devoted to formulate the problem using the BR and BRR measurements in the presence of $N_t$ transmitters and $N_r$ receivers. 
    Section III describes the radar uncertainty principle in mathematical expressions and formulates a non-convex optimization problem to improve the localization performance. In Section IV, numerical results are given to demonstrate the performance improvement of the proposed mechanism. Finally, Section V concludes the paper.

\section{Measurement Model and Cramer-Rao Lower Bound}\label{Preliminaries}
    Consider a distributed MIMO radar system which is consisting of $N_t$ moving or stationary Txs and $N_r$ moving or stationary Rxs in the three-dimensional (3-D) space. The position of Txs and Rxs are known and represented by 
    ${\bs{x}_{t,i}^{}=[x_{t,i}^{}\, y_{t,i}^{}\, z_{t,i}^{}]^T}$  and ${\bs{x}_{r,j}^{}=[x_{r,j}^{}\, y_{r,j}^{}\, z_{r,j}^{}]^T}$ 
    for ${i=1,\dots,N_t}$ and ${j=1,\dots,N_r}$, respectively. 
    Moreover, 
    ${\dot{\bs{x}}_{t,i}^{}=[\dot{x}_{t,i}^{}\, \dot{y}_{t,i}^{}\, \dot{z}_{t,i}^{}]^T}$  
    and 
    ${\dot{\bs{x}}_{r,j}^{}=[\dot{x}_{r,j}^{}\, \dot{y}_{r,j}^{}\, \dot{z}_{r,j}^{}]^T}$ 
    denote the velocity of Txs and Rxs for 
    ${i=1,\dots,N_t}$ and ${j=1,\dots,N_r}$, 
    respectively.
    Considering a moving target with position ${\bs{x}_{0}^{}=[x_{0}^{}\, y_{0}^{}\, z_{0}^{}]^T}$ and velocity ${\dot{\bs{x}}_{0}^{}=[\dot{x}_{0}^{}\, \dot{y}_{0}^{}\, \dot{z}_{0}^{}]^T}$, the true time delay 
    and the Doppler shift of the received reflected signal from the target at the $j$th Rx due to the $i$th Tx
    are denoted by
    \begin{equation}
        \setlength\arraycolsep{2.5pt}
        \begin{array}{l}
        t_{i,j}^{}\,=\cfrac{\,1\,}{c}\,({d_{t,i}^{}+d_{r,j}^{}}{})\\
        f_{i,j}^{}=\cfrac{f_c}{c}\,({\dot{d}_{t,i}^{}+\dot{d}_{r,j}^{}}{})
        \end{array},
    \end{equation}
    where $c$ and $f_c$ are the speed of light and carrier frequency, respectively.

    Moreover,
    $d_{t,i}^{}$ and $\dot{d}_{t,i}^{}$ denoting the range and range-rate between the target and the $i$th Tx,  are
    \begin{equation}
        \begin{array}{l}
        d_{t,i}^{}=\|\bs{x}_{t,i}^{}-\bs{x}_{0}^{}\|\\
        \dot{d}_{t,i}^{}=(\bs{x}_{t,i}^{}-\bs{x}_{0}^{})^T(\dot{\bs{x}}_{t,i}^{}-\dot{\bs{x}}_{0}^{})/d_{t,i}^{}
        \end{array}.
    \end{equation}
    Similarly, the range and range-rate between the $j$th Rx and the target are
    \begin{equation}
        \begin{array}{l}
        d_{r,j}^{}=\|\bs{x}_{r,j}^{}-\bs{x}_{0}^{}\|\\
        \dot{d}_{r,j}^{}=(\bs{x}_{r,j}^{}-\bs{x}_{0}^{})^T(\dot{\bs{x}}_{r,j}^{}-\dot{\bs{x}}_{0}^{})/d_{r,j}^{}
        \end{array}.
    \end{equation}

    Assuming common reference time for all Txs and Rxs, localization equations can be written generally as
    \begin{equation}\label{eq:bi-st}
        \begin{cases}
        d_{t,i}^{}+d_{r,j}^{}=r_{i,j}^{}\\
        {\dot{d}_{t,i}^{}+\dot{d}_{r,j}^{}}{}={\dot{r}_{i,j}^{}}{}
        \end{cases},
    \end{equation}
    in which $r_{i,j}^{}$ and $\dot{r}_{i,j}^{}$ denote the bi-static range and bi-static range rate data, respectively.
    In the presence of estimation error, measured BR and BRR can be modeled as
    \begin{equation}
        \begin{array}{l}
        \hat{\bs{r}}=\bs{r}+\Delta\bs{r}\\
        \hat{\dot{\bs{r}}}=\dot{\bs{r}}+\Delta\dot{\bs{r}}
        \end{array},
    \end{equation}
    where ${[\bs{r}]_{k}^{}=r_{i,j}^{}}$ and ${[\dot{\bs{r}}]_{k}^{}=\dot{r}_{i,j}^{}}$ for ${k=(i-1)N_{r}^{}+j}$. 
    In case the measurements are obtained via maximum likelihood estimators and the Txs' \color{\revisioncolor} signals occupy different frequency bands, \color{black}
    $[\Delta\bs{r}^T,\Delta\dot{\bs{r}}^T]^T$ can be modeled asymptotically (with respect to the received samples) as a zero-mean Gaussian random vector with covariance matrix \color{\revisioncolor}\cite{Wax1982}
    \begin{equation}
        \bs{\Sigma}^{}=\text{diag}
        \left(
        \begin{bmatrix}
        \bs{\sigma}\\
        \dot{\bs{\sigma}}
        \end{bmatrix}
        \right)
    \end{equation}
    where ${\bs{\sigma}{=}[{\sigma}_{1,1}^{2},\dots,{\sigma}_{N_t,N_r}^{2}]^T}$ and ${\dot{\bs{\sigma}}{=}[{\dot{\sigma}}_{1,1}^{2},\dots,{\dot{\sigma}}_{N_t,N_r}^{2}]^T}$. 
    For the interested parameter ${\theta}=(\bs{x}_0,\,\dot{\bs{x}}_0)$ and distribution ${\mathcal{N}(\bs{\mu}({\theta})\triangleq[\bs{r}^{T},\dot{\bs{r}}^{T}]^T,\,\bs{\Sigma})}$, the Fisher information matrix is given by \cite{kay1993fundamentals}
    \color{black}
    \begin{equation}
        \fim
        =\left[
        \frac{\partial \bs{\mu}}{\partial \bs{x}_0}\  \frac{\partial \bs{\mu}}{\partial \dot{\bs{x}}_0}
        \right]^T
        \bs{\Sigma}^{-1}
        \left[
        \frac{\partial \bs{\mu}}{\partial \bs{x}_0}\  \frac{\partial \bs{\mu}}{\partial \dot{\bs{x}}_0}
        \right]
    \end{equation}
    where
    \begin{equation}
        \cfrac{\partial \bs{\mu}}{\partial \bs{x}_0}=
        \begin{bmatrix}
            \cfrac{\partial \bs{r}}{\partial \bs{x}_0}
            \\
            \cfrac{\partial \dot{\bs{r}}}{\partial \bs{x}_0}
        \end{bmatrix}_{2N_{t}^{}N_{r}^{}\times 3},\quad
        \cfrac{\partial \bs{\mu}}{\partial \dot{\bs{x}}_0}=
        \begin{bmatrix}
            \cfrac{\partial \bs{r}}{\partial \dot{\bs{x}}_0}
            \\
            \cfrac{\partial \dot{\bs{r}}}{\partial \dot{\bs{x}}_0}
        \end{bmatrix}_{2N_{t}^{}N_{r}^{}\times 3}.
    \end{equation}
    By differentiating \eqref{eq:bi-st} with respect to $\bs{x}_0$ and $\dot{\bs{x}}_0$ we have
    \begin{gather}
        \label{eq:grad}
        \begin{array}{lll}
        \bs{\rho}_{k}^{\color{black}T\color{black}}\triangleq
        \left[\cfrac{\partial \bs{r}}{\partial \bs{x}_0}\right]_{k,:}^{}
        =
        \bs{\rho}_{t,i}^T + \bs{\rho}_{r,j}^T
        ,&&
        \cfrac{\partial \bs{r}}{\partial \dot{\bs{x}}_0}=\mathbf{O},
        \\
        \dot{\bs{\rho}}_{k}^{\color{black}T\color{black}}\triangleq
        \left[\cfrac{\partial \dot{\bs{r}}}{\partial \bs{x}_0}\right]_{k,:}^{}=
        \dot{\bs{\rho}}_{t,i}^T + \dot{\bs{\rho}}_{r,j}^T
        ,&&
        \cfrac{\partial \dot{\bs{r}}}{\partial \dot{\bs{x}}_0}=\cfrac{\partial {\bs{r}}}{\partial {\bs{x}}_0},
        \end{array}
    \end{gather}
    where
    \begin{gather}\label{eq:rho}
        \begin{array}{l}
        \bs{\rho}_{t,i}^{}=\cfrac{\partial d_{t,i}^{}}{\partial {\bs{x}}_0^{}}=\cfrac{{\bs{x}}_0^{}{-}\bs{x}_{t,i}^{}}{d_{t,i}^{}},\\
        \dot{\bs{\rho}}_{t,i}^{}=\cfrac{\partial \dot{d}_{t,i}^{}}{\partial {\bs{x}}_0^{}}=\cfrac{\dot{\bs{x}}_0^{}{-}\dot{\bs{x}}_{t,i}^{}{-}\dot{d}_{t,i}^{}\bs{\rho}_{t,i}^{}}{d_{t,i}^{}}.
        \end{array}
    \end{gather}
    Similarly, $\bs{\rho}_{r,j}^{}$ and $\dot{\bs{\rho}}_{r,j}^{}$ are also computed by replacing subscript $(t,i)$ with $(r,j)$ in \eqref{eq:rho}.
    Finally, the CRLB matrix of the vector parameter $[\bs{x}_0^T,\ \dot{\bs{x}}_0^T]^T$ can be written as 
    \begin{equation}
    \crb=\fim^{-1}.
    \end{equation}

\section{Radar Uncertainty Principle and Conceptual Waveform Design}
    \subsection{Radar Uncertainty Principle}
    For a radar signal $s(t)$, the TD accuracy ($\delta \tau$) and DS accuracy ($\delta f$) are expressed as 
    \begin{equation}
    \delta \tau = \frac{1}{B_{\text{eff}}\sqrt{2E/N_0}},\quad
    \delta f = \frac{1}{T_{\text{eff}}\sqrt{2E/N_0}}
    \end{equation}
    where $E$ is the signal energy, $N_0$ is the noise power per unit bandwidth, $T_{\text{eff}}$ and $B_{\text{eff}}$ are the \emph{effective time duration} and \emph{effective bandwidth} of 
    $s(t)$ respectively\cite{Merrill2001}.
    For a given time-bandwidth product (compression ratio),
    radar uncertainty principle indicates that 
    "there is an inherent invariance in the product of the
    range and range rate measurement accuracy and resolution; by changing the signal form, it is possible to change the accuracy of the range and range rate measurement in such a manner that a gain for one parameter leads to a loss for the other" \cite{Vakmann1968}.
    Therefore,
    waveform design at each Tx is restricted to strike a balance between time and bandwidth which leads to a trade-off between time-delay and Doppler-shift measurement accuracy. 
    In the same spirit, radar uncertainty principle in distributed MIMO radars can be expressed by the following mathematical expressions as 
    \begin{equation}\label{eq:uncertainty}
    {\sigma}_{i,j}^{}{\dot{\sigma}}_{i,j}^{} = C_{i,j}^{}, \quad \cfrac{{\sigma}_{i,j}^{}}{{\dot{\sigma}}_{i,j}^{}}=\cfrac{{\sigma}_{i,j'}^{}}{{\dot{\sigma}}_{i,j'}^{}}={\alpha_{i}^{}},
    \end{equation}
    in which $j'=1,\dots,N_r$ and the value of $C_{i,j}^{}$ is related to \mbox{1) the} 
    \color{\revisioncolor} ratio of the $i$th Tx's signal energy to  the noise power at the $j$th Rx, and 2) the given time-bandwidth product for the $i$th Tx. In addition, $\alpha_{i}^{}$ quantifies the trade-off between accuracy the TD and DS measurements. 
    The first identity expresses the uncertainty principle for the pair $(\text{Tx}_{i}^{},\,\text{Rx}_{j})$ and the second one
    \color{black} indicates that the waveform at the $i$th Tx has a same proportional effect $\alpha_{i}^{}$ on the BR and BRR measurements accuracy for all Rxs. 
    Thus, the proportional coefficient $\alpha_{i}^{}$ is considered as the design parameter at the $i$th Tx and can be adjusted to improve the MIMO radar performance in a manner. 
    In other words, we determine that each Tx provides high accuracy in range or velocity estimation in order to have optimum MIMO radar performance.
    \subsection{Conceptual Design}
    Defining 
    ${\bs{c}{=}[C_{1,1}^{},\dots,C_{N_t,N_r}^{}]^T}$ and 
    ${\bs{\alpha}{=}[\alpha_{1}^{},\dots,\alpha_{N_t}^{}]^T}$,
    we write the following optimization problem to improve the \emph{weighted} localization performance at the estimated target position $\hat{\bs{x}}_{0}^{}$ and velocity $\hat{\dot{\bs{x}}}_{0}^{}$
    \begin{equation}\label{eq:trace}
        \begin{array}{cl}
        \min\limits_{\bs{\sigma},\dot{\bs{\sigma}},\bs{\alpha}} & 
        \tr\left\{\bs{W}\times\crb|_{\hat{\bs{x}}_{0}^{},\hat{\dot{\bs{x}}}_{0}^{}}\right\}\\
        \text{s.t.} & 
        \sigma_{k}^{}\dot{{\sigma}}_{k}^{}=c_{k}^{2}\\
        &
        {c}_{k}^{} \alpha_{i}^{}\,= \sigma_{k}^{},\\
        &
        l_{i}^{}\leq\alpha_{i}^{}\leq u_{i}^{},
        \end{array}
    \end{equation}
    where, for simplicity of notation, the $k$th element of a vector $\bs{a}$ is denoted by $a_{k}^{}$; and $\bs{W}$ is a positive definite weight matrix. Moreover, ${l}_{i}^{}$ and ${u}_{i}^{}$ are the lower and upper bound for the parameter design at the $i$th Tx respectively, which are induced by practical limitations.
    Furthermore, the target position and velocity can be estimated using methods in \cite{yang2016improved,amiri2017efficient,Amiri2017,amiri2017asymptotically}.
    The constrained optimization problem in \eqref{eq:trace} can be converted to a bound constrained problem by substituting 
    \begin{equation}
        \sigma_{k}^{}={c_{k}^{}}{\alpha_{i}^{}},\quad
        \dot{\sigma}_{k}^{}=\cfrac{c_{k}^{}}{\alpha_{i}^{}}
    \end{equation}
    into the objective function and performing some manipulations. That is, \eqref{eq:trace} takes the following form as
    \begin{equation}\label{eq:bound}
        \begin{array}{cl}
        \min\limits_{\bs{\alpha}}^{}&
        f(\bs{\alpha})=\tr
        \left\{ 
        \bs{W}
        \left[
        \displaystyle\sum\limits_{i=1}^{N_{t}^{}}
        \left(
        \frac{1}{\alpha_{i}^{}}\,\bs{P}_{i}^{}
        +
        \alpha_{i}^{}{\bs{V}}_{i}^{}
        \right)
        \right]^{-1}
        \right\}\\
        \text{s.t.}
        &
        l_{i}^{}\leq\alpha_{i}^{}\leq u_{i}^{},
        \end{array}
    \end{equation}
    in which the term
    $\displaystyle\sum\limits_{i}^{}
    \big(
    \frac{1}{\alpha_{i}^{}}\,\bs{P}_{i}^{}+\alpha_{i}^{}{\bs{V}}_{i}^{}
    \big)$
    is the new representation of $\fim$,
    \begin{align}
        \bs{P}_{i}^{}&=
        \sum_{j=1}^{N_{r}^{}}c_{k}^{-1}
        \begin{bmatrix}
        \bs{\rho}_{k}^{}\\
        \bs{0}
        \end{bmatrix}
        \begin{bmatrix}
        \bs{\rho}_{k}^{T}&
        \bs{0}^T
        \end{bmatrix}
        ,\nonumber\\
        {\bs{V}}_{i}^{}&=
        \sum_{j=1}^{N_{r}^{}}c_{k}^{-1}
        \begin{bmatrix}
        \dot{\bs{\rho}}_{k}^{}\\
        {\bs{\rho}}_{k}^{}
        \end{bmatrix}
        \begin{bmatrix}
        \dot{\bs{\rho}}_{k}^{T}&
        {\bs{\rho}}_{k}^{T}
        \end{bmatrix},
    \end{align}
    and recall that ${k=(i-1)N_{r}^{}+j}$.
    The $i$th element of the gradient vector is also determined by \cite{petersen2008matrix}
    \begin{equation}\label{eq:grad}
    g_{i}^{}\triangleq
    \cfrac{\partial f}{\partial \alpha_{i}^{}}=\tr
    \left\{
    \left(\cfrac{\partial f}{\partial \fim}\right)^T \cfrac{\partial \fim}{\partial \alpha_{i}^{}}
    \right\}
    \end{equation}
    where
    \begin{align}
    \cfrac{\partial f}{\partial \fim}=-\fim^{-1}\bs{W}\fim^{-1},\quad
    \cfrac{\partial \fim}{\partial \alpha_{i}^{}}=\bs{V}_{i}^{}-\frac{1}{\alpha_{i}^{2}}\,\bs{P}_{i}^{}.
    \end{align}
    \color{\revisioncolor}
    Here, iterative algorithms such as \emph{sequential quadratic programming} (SQP) with BFGS Hessian approximation can be utilized to find a local solution to the bound constrained optimization problem in \eqref{eq:bound}\cite{nocedal2006numerical}. 
    \color{black}Heuristics algorithms such as \emph{particle swarm} can also be applied to obtain the global solution.

\section{Numerical Results}
\begin{figure}[!t]
    \centering
    \includegraphics{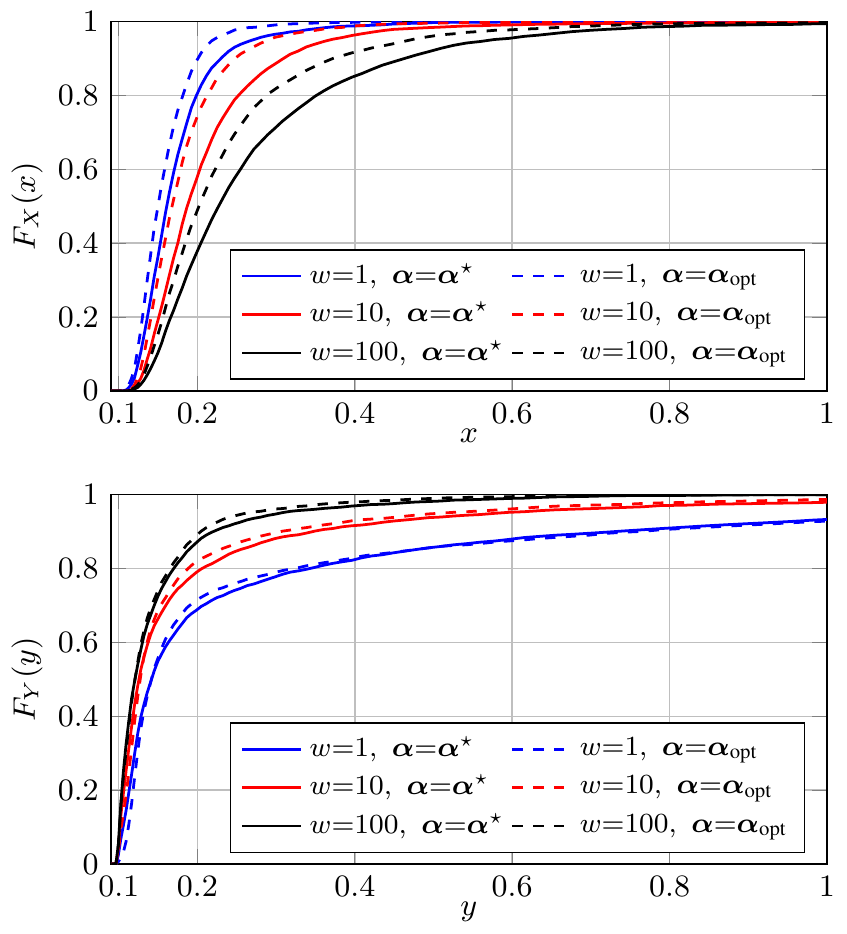}
    \vspace{-10pt}
    \caption{Cumulative distribution function of the CRLB improvement ratio for the target position (top) and target velocity (bottom),
    $\bs{\alpha}^{\star}$ and $\bs{\alpha}_{\text{opt}}$ and three values of $w$.}
    \label{fig2}
    \end{figure}

    The following simulation are carried out in order to
    evaluate the performance of the proposed conceptual waveform design.
    The noise in the measured BR and BRR
    are modeled similar to \cite{noroozi2016weighted} and \cite{amiri2017asymptotically} 
    as zero-mean Gaussian random variables 
    which their standard deviations product dependent only on the signal-to-noise ratio at each pair Tx-Rx. 
    Therefore, 
    the BR and BRR measurements were corrupted by additive Gaussian noise with product of the standard deviations
    ${{{\sigma _{i,j}}\dot{\sigma} _{i,j}} = ({\sigma _0}{d_{t,i}}{d_{r,j}})^2/R^4}$
    for ${i = 1, \dots ,N_t}$ and  ${j = 1, \dots ,N_r}$, where ${{\sigma _0}}$ is a constant and
    $R$ denotes the radius of the surveillance area which is equal to
    $6000$ m (see \cite{amiri2017asymptotically} and \cite{noroozi2016weighted} for further details).
    We also set ${l_{i}=l=1}$ and ${u_{i}=u=100}$ for $i=1,\dots,N_{t}$. 
    The weight matrix takes the following form in all simulations.
    \begin{equation*}
    \bs{W}=
    \begin{bmatrix}
        \bs{I}_3 & \mathbf{O}\\
        \mathbf{O} & w\bs{I}_3
    \end{bmatrix}
    \end{equation*}
    Without loss of functionality, the objective function in \eqref{eq:bound} is computed at the true values of the target position and velocity. 
    We consider 5000 configurations of a distributed MIMO radar system with four Txs and six Rxs which are located randomly within the 3-D region with 
    $({R/2\leq\rho\leq R};\  {0\leq\phi\leq 2\pi};\  {200\leq z \leq 300 \text{m}})$
    in cylindrical coordinate.
    The target position is also realized randomly within the 3-D region with 
    $({0\leq\rho\leq 2R};\  {0\leq\phi\leq 2\pi};\  {300\leq z \leq 600\  \text{m}})$
    in cylindrical coordinate. 
    The velocities of the Txs, Rxs, and target are set to random velocities whose maximum magnitude is $100$ m/s.
    Using MATLAB Optimization Toolbox \cite{MatlabOTB}, a 
    local solution to the optimization problem in \eqref{eq:bound} is obtained via the SQP 
    algorithm with initial point $\bs{\alpha}_{0}=\sqrt{lu}{\times}\bs{1}_{N_t}$ and denoted by $\bs{\alpha}^{\star}$. 
    Moreover, $\bs{\alpha}_{\text{opt}}$ which denotes the global solution to the problem, is obtained by particle swarm algorithm. 
    CRLB is evaluated at $\bs{\alpha}^{\star}$, $\bs{\alpha}_{\text{opt}}$ and $\bs{\alpha}_{0}$ for all configurations.
    Cumulative distribution functions (CDF) of the random variables $X$ and $Y$,
    \begin{equation*}
    X{=}\frac
    {\displaystyle\sum_{k=1}^{3}\big[\crb(\bs{\alpha})\big]_{k,k}}
    {\displaystyle\sum_{k=1}^{3}\big[\crb(\bs{\alpha}_{0})\big]_{k,k}}, \quad
    Y{=}\frac
    {\displaystyle\sum_{k=4}^{6}\big[\crb(\bs{\alpha})\big]_{k,k}}
    {\displaystyle\sum_{k=4}^{6}\big[\crb(\bs{\alpha}_{0})\big]_{k,k}},
    \end{equation*}
    which are the CRLB improvement ratios for the target position and velocity respectively, are shown in Fig. \ref{fig2} for $\bs{\alpha}^{\star}$ and $\bs{\alpha}_{\text{opt}}$ and three values of $w$. 
    The figure confirms a significant improvement ratio for the target position and velocity using the proposed mechanism.
    As expected, higher value of $w$ leads to higher CRLB improvement ratio of the target velocity and vice versa. 
    According to the figure, it is also concluded that the global solution has more impact on the CRLB improvement ratio of the target position with respect to the target velocity. 
    Hence, for the scenarios which require high accuracy for the target position,
    the global solution is recommended regardless of its higher computational complexity.  

\begin{figure}[!t]
    \centering
    \includegraphics{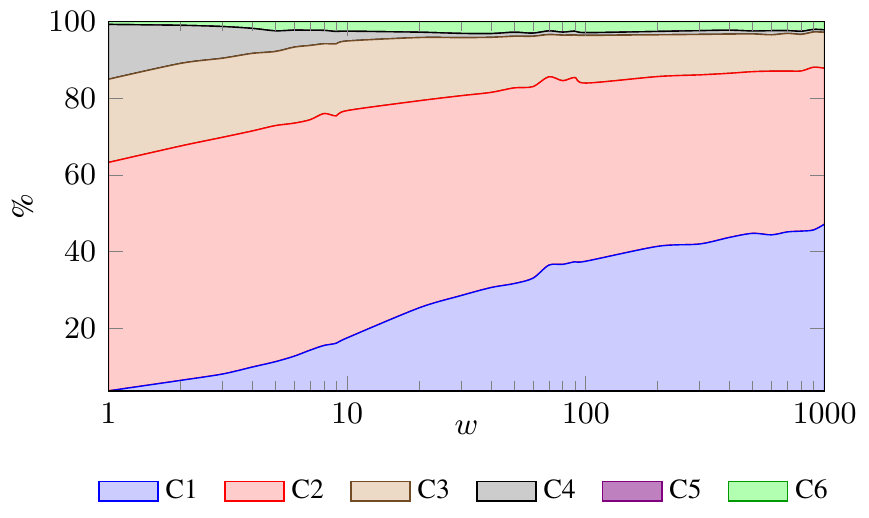}
    \vspace{-10pt}
    \caption{Percentage of the configurations that $\bs{\alpha}_{\text{opt}}$ belong to each cluster C1 to C6 for different values of $w$.}
    \label{fig3}
    \end{figure}

    We also show that vertices of the bounding box are appropriate candidates for the global solution to \eqref{eq:bound}.
    To this end, the global solutions are classified into the following six clusters
    \begin{enumerate}[label=C\arabic*:]
        \item 
        $\bs{\alpha}_{\text{opt}}=[u,u,u,u]^T$ (i.e. accurate DS for all Txs),
        \item 
        $\bs{\alpha}_{\text{opt}}$ is a permutation of $[l,u,u,u]^T$,
        \item 
        $\bs{\alpha}_{\text{opt}}$ is a permutation of $[l,l,u,u]^T$,
        \item 
        $\bs{\alpha}_{\text{opt}}$ is a permutation of $[l,l,l,u]^T$,
        \item 
        $\bs{\alpha}_{\text{opt}}=[l,l,l,l]^T$,
        \item 
        $\bs{\alpha}_{\text{opt}}$ is not a vertex.
    \end{enumerate}
    According to the Fig. \ref{fig3}, which illustrates percentage of the configurations belong to each cluster for different values of $w$, it results that a vertex of the bounding box is the global solution for almost all configurations. Since $N_t$-dimensional box has $2^{N_t}$ vertices, evaluating and comparing the objective function at all vertices is an admissible method (in terms of computational complexity) to find the global solution for the small number of Txs (namely $N_t\leq 10$). Whereas, for the large number of Txs, a local solution can be obtained by lower computational complexity using iterative algorithms.
    Fig. \ref{fig3} also indicates that the DS accuracy is generally more important than the TD for improving the CRLB in a distributed MIMO radar. As we see in the figure, $63\%$ of all configurations take the optimum weighted CRLB where at least three Txs facilitated to provide high accuracy for DS measurements.

\section{Conclusions}
    In this paper, the problem of localizing a moving target in
    the 3-D space from TD and DS information gathered from a
    widely distributed MIMO radar system was considered.
    A conceptual waveform design for transmitters was proposed based on the radar uncertainty principle in order to improve the CRLB. Numerical results were also included to illustrate the effectiveness of the proposed mechanism on the CRLB of the target position and velocity. By numerical results, we also concluded that the DS accuracy has more impact on the CRLB improvement. 
    In future works, it will be interesting to propose mathematical approvement for the consequences which are taken from the numerical results

\bibliographystyle{IEEEtran}
\bibliography{main}
\end{document}